\documentclass[NumberedRefs,turnofflinenumbers]{JASA-EL}
\usepackage{amsmath}
\usepackage{amsthm}
 \usepackage{graphicx}
  \usepackage{color}
\usepackage{amssymb, latexsym}

\usepackage[tight,footnotesize]{subfigure}

\begin{document}

\title[Three dimensional higher-order raypath separation]{Three dimensional higher-order raypath separation
in a shallow-water waveguide}


 \author{Jiang Longyu}
 \email{JLY@seu.edu.cn}
 \correspondingauthor
 \author{Zhang Zhe}
 \email{220151575@seu.edu.cn}
  \affiliation{The Laboratory of Image Science and Technology, Southeast University, Nanjing 210096, China, Centre de Recherche en Information Biomedicale Sino-Francais (LIA CRIBs), Rennes, France.}

  \author{Roux Philippe}
 \email{philippe.roux@univ-grenoble-alpes.fr}
 \affiliation{{Institut des Sciences de la Terre, Universit\'{e} Joseph Fourier, Centre National de la Recherche Scientifique,1381 Rue de la Piscine, Saint-Martin d'H\`{e}res, France.}}



\date{\today}

\begin{abstract}
Separating raypaths in a multipath shallow-water environment is a challenge problem due to the interferences between them and colored noise existing in ocean environment, especially for two raypaths arrive close to each other. Thus, in this paper, a three dimensional (3D) higher-order raypath separation in an array to array configuration is proposed. Performance tests using simulation data in a multipath environment, real data obtained in an ultrasonic waveguide and ocean shallow-water data, respectively, illustrate that the proposed algorithm achieves a higher resolution and a stronger robustness comparing to the existing algorithms.
\end{abstract}


\maketitle

\section{Introduction}

Acoustic rays are multi-path propagation in shallow-water due to the reflection and (or) the refraction by the surface of the ocean and the continental shelf. The multiple raypaths cover more parts of the ocean and provide more information than a single emitted signal. However, they also produce interferences among them. Thus, separating the raypaths is necessary in many important applications, such as passive sonar, ocean acoustic tomography, etc.
As a classical separation algorithm, the Multiple Signal Classification (MUSIC) algorithm \cite{schimidt1986} was proposed to obtain a high-resolution separation of the sources in a point to array configuration, which is composed of a point source and a vertical receiver array. The MUSIC algorithm improves separation resolution mainly exploiting the orthogonality between the signal subspace and the noise subspace. However, it fails when the signals are fully correlated or coherent. Jiang et el. \cite{jiang2017raypath} proposed a smoothing active wideband MUSIC (Smoothing-MUSICAL) algorithm, which is an extension of the MUSIC algorithm to the case of separating fully correlated or coherent wideband signals by using the spatial-frequency smoothing. In addition, comparing to the MUSIC algorithm, the smoothing-MUSICAL algorithm improves the separation resolution through adding the arrival time of each raypath to the signal model as a new discrimination parameter.  Recently, a  raypath separation algorithm based on the fourth order cumulant \cite{jiang2017} is proposed in a point to array configuration. Higher resolution and more robust separation is obtained due to the properties of aperture extension and noise suppression of fourth-order cumulants while at a cost of computation time. Sequentially, Jiang et el. propose a fast algorithm using low-rank matrix approximation \cite{Jiang2018Fast} to reduce the computation cost of the 4-smoothing-MUSICAL algorithm. On the other hand, Roux et el \cite{Roux2008The}   develop an array to array configuration, which is composed of a source vertical array and a receiver vertical array.  A double-beamforming algorithm is introduced experimentally in the configuration and a high-resolution tomography inversion \cite{Iturbe2008Ocean} is finally obtained through using its separation results. Moreover, Touz\'{e} et el \cite{Touze2012Double} present the double-capon algorithm and the double-MUSICAL algorithm under the assumption of white Gaussian noise. To further improve both the resolution and the robustness to the colored noise, we present a 3D higher-order raypath separation algorithm in an array to array configuration in this paper. Its performance improvement is achieved due to inheriting the merits of using both the higher-order cumulants and the array to array configuration.

The rest of the paper is organized as follows.  In Section 2, we give an elaborate description of the 3D higher-order raypath separation algorithm. In Section 3, we test the performance of the proposed algorithm using simulation data in a multi-path environment, real data obtained in an ultrasonic tank and ocean data, respectively. We conclude the paper in Section 4.

 \section{The 3D higher-order raypath separation algorithm}

 \subsection{Signal model}

   Assume the raypaths propagate in a double vertical array configuration, which is composed of an emission array ($N$ sources) and a receiver array ($M$ sensors).  The $P$ raypaths at the frequency $\nu$ produced by the $n^{th}$ source and received on the $m^{th}$ receiver is noted as $x_{m,n,\nu}$ and modeled as follows.

 \begin{equation}
 \centering
 x_{m,n,\nu}=s_{\nu}\sum_{p=1}^{P} a_p e^{\Phi_p}+b_{m,n,\nu}
 \end{equation}
 where $s_{\nu}$ is the source spectrum, $\Phi_p=-j2\pi\nu (T_p+(m-m_0)\tau_{p}^{e}+(n-n_0)\tau_{p}^{r})$, $\tau_{p}^{e}=d sin(\theta_{p}^{e})/c$ and $\tau_{p}^{r}=d sin(\theta_{p}^{r})/c$. $T_p$ notes the arrival time of the $p^{th}$ raypath. $m_0$ ($n_0$) refers to the reference source (receiver). $\tau_{p}^{e}$ ($\tau_{p}^{r}$) is the time delay for the $p^{th}$ raypath propagates between two adjacent sources (receivers). $d$ is the interval between two adjacent sources (receivers). $ \theta_{p}^{e}$ ($\theta_{p}^{r}$) is the direction of emission (reception) of the $p^{th}$ raypath. $a_p$ notes the amplitude of the $p^{th}$ raypath. $b_{m,n,\nu}$ refers to the additive noise.

 \subsection{Data model}

Based on the signal model built in above section, each signal received in the double-array configuration is a data cube for all $N$ sources, $M$ sensors and $F$ frequencies of the wideband signals considered in the algorithm. The data model is built through concatenating its elements into a long vector.
First, the elements corresponding to the source $n^{th}$ at the frequency $\nu$ on all the $M$ elements of the receiver array is concatenated into the vector as follows ( $+$ notes transpose.) :
 \begin{equation}
  \centering
{\bf {x}}_{n,\nu}= [ x_{1,n,\nu} \ldots x_{M,n,\nu}]^{+}
 \end{equation}

The elements corresponding to both the $N$ source and the $M$ sensors are further concatenated into the following vector.
 \begin{equation}
  \centering
{\bf {x}}_{\nu}= [ {\bf {x}}_{1,\nu} \ldots {\bf {x}}_{n,\nu}]^{+}
 \end{equation}
Finally, considering all the $F$ frequencies, the long vector ${\bf {X}}$ is obtained as follows.
 \begin{equation}
  \centering
{\bf {X}}= [ {\bf {x}}_{\nu_1} \ldots {\bf {x}}_{\nu_F}]^{+}
 \end{equation}

Similarly, the steering vector $\mathbf d( \mathbf \theta_{p}^{e} \mathbf \theta_{p}^{r}, \mathbf T_p )$ and the additive noise $\mathbf b$ can be also concatenated into a long vector.
Thus, the received signal in the frequency domain can be written in the following matrix form.

 \begin{equation}
 \centering
 \mathbf X=\sum_{p=1}^{P} a_p \mathbf d( \mathbf \theta_{p}^{e}, \mathbf \theta_{p}^{r}, \mathbf T_p )+\mathbf b= \mathbf D(\mathbf \theta_{p}^{e}, \mathbf \theta_{p}^{r}, \mathbf T_p)+\mathbf b
 \end{equation}
where

 $\mathbf X=[ \mathbf x_{\nu_1}, \mathbf x_{\nu_2},\cdots, \mathbf x_{\nu_F}]^{+}$, $ \mathbf x_{\nu}=[\mathbf x_{1, \nu}, \mathbf x_{2, \nu}, \cdots, \mathbf x_{N, \nu}] ^{+} \ \ (\nu= \nu_1, \nu_2, \cdots, \nu_F) $, and $\mathbf x_{n,\nu}=[x_{1,n,\nu}, x_{2,n,\nu}, \cdots, x_{M,n,\nu}]^{+} \ (n=1, 2,\cdots, N)$. $\mathbf \theta_{p}^{e} =[ \theta_{1}^{e},  \theta_{2}^{e}, \cdots, \theta_{P}^{e}]^{+} $, $\mathbf \theta_{p}^{r} =[\theta_{1}^{r},   \theta_{2}^{r}, \cdots, \theta_{P}^{r}]^{+} $, $\mathbf T_p=[T_1,  T_2, \cdots, T_P]^{+} $.

 $\mathbf d( \mathbf \theta_{p}^{e}, \mathbf \theta_{p}^{r}, \mathbf T_p )=[\mathbf d_{\nu_1}( \mathbf \theta_{p}^{e}, \mathbf \theta_{p}^{r}, \mathbf T_p ), \mathbf d_{\nu_2}( \mathbf \theta_{p}^{e}, \mathbf \theta_{p}^{r}, \mathbf T_p ), \cdots, \mathbf d_{\nu_F}( \mathbf \theta_{p}^{e}, \mathbf \theta_{p}^{r}, \mathbf T_p )]^{+}$, with $\mathbf d_{\nu_i}(\mathbf \theta_{p}^{e}, \mathbf \theta_{p}^{r}, \mathbf T_p )=[s_{\nu_i}e^{-j2\pi\nu_i ( T_p+ (1-n_0)\tau(\theta_{p}^{e}) )}\mathbf d (\mathbf \theta_{p}^{r})_{\nu_i}, \cdots s_{\nu_i}e^{-j2\pi\nu_i (T_p+ (N-n_0)\tau(\theta_{p}^{e}))}\mathbf d (\mathbf \theta_{p}^{r})_{\nu_i} ]^{+}$ ($i=1, 2, \dots, F$) and
$\mathbf d (\mathbf \theta_{p}^{r})_{\nu_i}=[e^{-j2\pi\nu_i (1-m_0)\tau(\theta_{p}^{r})}, \cdots, e^{-j2\pi\nu_i (M-m_0)\tau(\theta_{p}^{r})}]^{+}$.

 $\mathbf b=[ \mathbf b_{\nu_1}, \mathbf b_{\nu_2},\cdots, \mathbf b_{\nu_F}]^{+}$, with $ \mathbf b_{\nu}=[\mathbf b_{1, \nu}, \mathbf b_{2, \nu}, \cdots, \mathbf b_{N, \nu}] ^{+} \ \ (\nu= \nu_1, \nu_2, \cdots, \nu_F) $, and $\mathbf b_{n,\nu}=[b_{1,n,\nu}, b_{2,n,\nu}, \cdots, b_{M,n,\nu}]^{+} \ (n=1, 2,\cdots, N)$

\subsection{Principle of the Algorithm}

Using the data model built above, the trispectrum matrix of the received signal can be theoretically computed according to the following equation.

\begin{equation}
\begin{array}{rcl}
\mathbf C=E\{(\mathbf X \bigotimes \mathbf X^{*})(\mathbf  X \bigotimes\mathbf X^{*})^{H}\}
\\-E\{(\mathbf X\bigotimes\mathbf X^{*}){\}}E\{(\mathbf X\bigotimes\mathbf X^{*})^{H}\}  \\
 -E\{(\mathbf X \mathbf X^{H})\}\bigotimes E\{(\mathbf X \mathbf X^{H})^{*}\},
 \end{array}
\label{eq:x5}
\end{equation}
However, in a real ocean experiment, it is costly to obtain several realizations to compute the expectation of the stochastic process. Moreover, the raypaths are reflected or refracted by the emitted signal thus they are fully correlated or coherent, which will lead to a rank deficiency of the trispectrum matrix. To resolve the problems, we use a three dimensional smoothing technique \cite{Touze2012Double} to generate several realizations based on the received signal. The smoothing is performed in the dimensions of the emitted array, the receiver array and the frequency bins, respectively. That is, the emitted (receiver) array is divided into $K_e$ ($K_r$) subarrays for a length of $N_{e}^{s}=N-K_e+1$ ($N_{r}^{s}=M-K_r+1$). Similarly the $K_f$  frequency subbands are respectively composed of $N_{f}^{s}=F-K_f+1$ frequency bins. Through these operation, we generate $K_e \times K_r \times K_f $  small data cubes, whose dimensions are $N_{e}^{s}\times N_{r}^{s} \times N_{f}^{s}$. The $\mathbf X_{i, j, k}$ is obtained through connecting the elements of all the three dimensions into a long vector. To be easily understood, the schematic diagram of the smoothing technique for the cumulants is provided in Fig. 1.

\begin{figure}
\centering     
{\includegraphics[width=120mm]{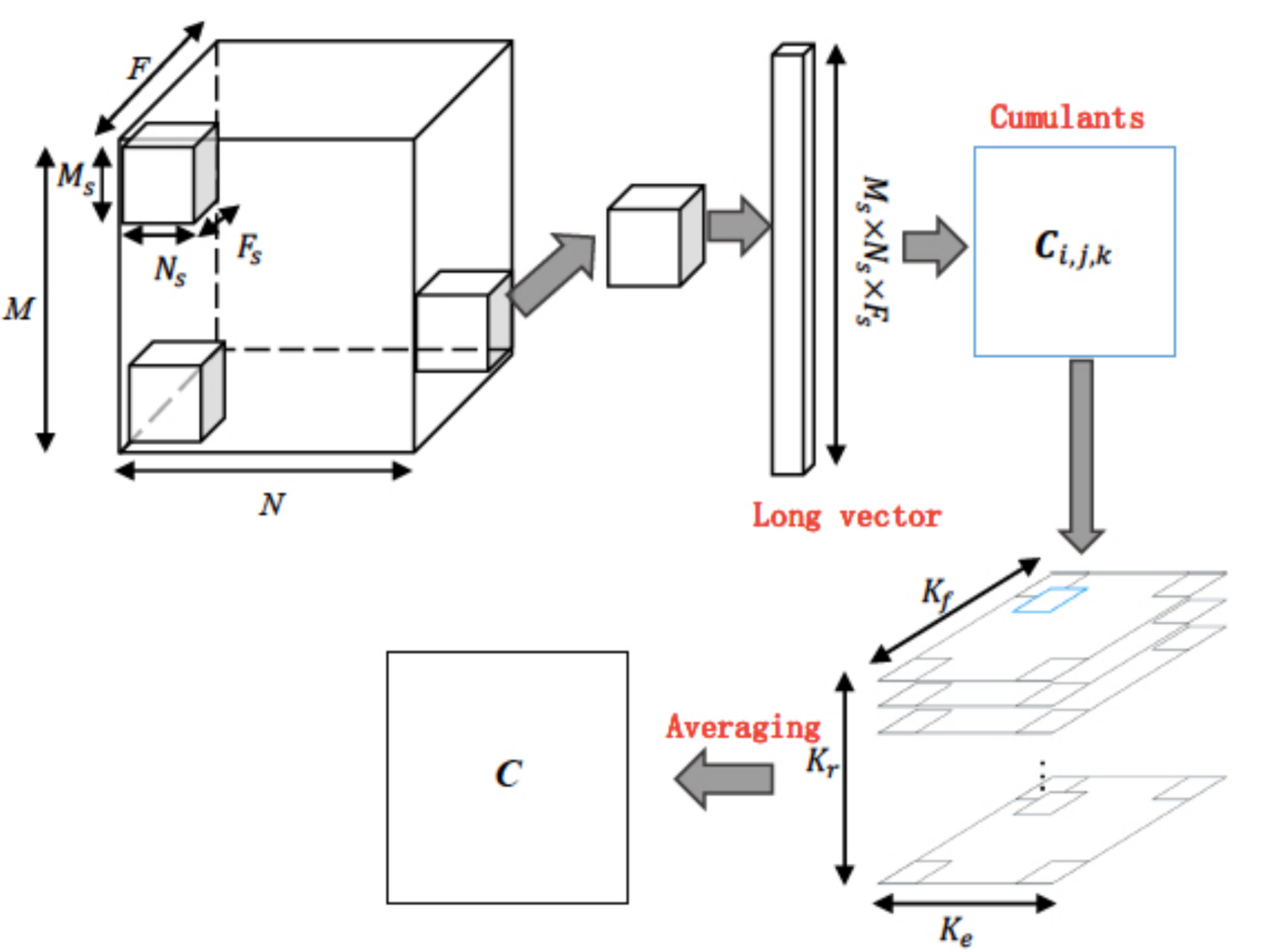}}
\caption{Schematic diagram of the smoothing technique for the cumulants.}
\end{figure}

The expectation of the trispectrum matrix of the received signal is finally estimated through an average of these trispectrum matrix computed using the small data cubes, which is specifically described in Equation 4.

\begin{equation}
\begin{array}{rcl}
 \widehat {\mathbf C}=\frac{1}{{{K}_{e}}{{K}_{r}}{{K}_{f}}}\sum\limits_{k=1}^{{{K}_{f}}}{\sum\limits_{j=1}^{{{K}_{e}}}{\sum\limits_{i=1}^{{{K}_{r}}}{{{\mathbf C}_{i,j,k}}}}} =\frac{1}{{{K}_{e}}{{K}_{r}}{{K}_{f}}}\sum\limits_{k=1}^{{{K}_{f}}}{\sum\limits_{j=1}^{{{K}_{e}}}}{\sum\limits_{i=1}^{{{K}_{r}}}} E\{(\mathbf X_{i, j, k} \bigotimes \mathbf X_{i, j, k}^{*})(\mathbf  X_{i, j, k} \bigotimes\mathbf X_{i, j, k}^{*})^{H}\}
\\-E\{(\mathbf X_{i, j, k}\bigotimes\mathbf X_{i, j, k}^{*}){\}}E\{(\mathbf X_{i, j, k}\bigotimes\mathbf X_{i, j, k}^{*})^{H}\}
 -E\{(\mathbf X_{i, j, k} \mathbf X_{i, j, k}^{H})\}\bigotimes E\{(\mathbf X_{i, j, k} \mathbf X_{i, j, k}^{H})^{*}\},
 \end{array}
\label{eq:x5}
\end{equation}

%
Apply eigenvalue decomposition to $\widehat {\mathbf C}$
\begin{equation}
\centering
\widehat {\mathbf C}=\mathbf U \mathbf \Lambda {{\mathbf U}^{H}}=\sum\limits_{i=1}^{{{({{N}_{s}}{{M}_{s}}{{F}_{s}})}^{2}}}{{{\lambda }_{i}}{{\mathbf u}_{i}} \mathbf u_{i}^{H}=}\sum\limits_{i=1}^{{{P}^{2}}}{{{\lambda }_{i}}{{\mathbf u}_{i}} \mathbf u_{i}^{H}}+\sum\limits_{i={{P}^{2}}+1}^{{{({{N}_{s}}{{M}_{s}}{{F}_{s}})}^{2}}}{{{\lambda }_{i}}{{\mathbf u}_{i}} \mathbf u_{i}^{H}}
\end{equation}
where $\mathbf U$ is composed of ($N_{e}^{s}\times N_{r}^{s} \times N_{f}^{s})^2$ eigenvectors and $ \mathbf \Lambda$
is a diagonal matrix containing the ($N_{e}^{s}\times N_{r}^{s} \times N_{f}^{s})^2$ eigenvalues.
$\mathbf u_{i} $ denotes the $i_{th}$ eigenvector. Arranging all the eigenvalues from largest to smallest $\lambda_1> \lambda_2 > \cdots,>\lambda{_{(N_{e}^{s}\times N_{r}^{s} \times N_{f}^{s})}}^2 $, the $(N_{e}^{s}\times N_{r}^{s} \times N_{f}^{s})^2-P^{2}$ eigenvectors corresponding to the smallest eigenvalues project the noise subspace $ {\mathbf U}_{n} \mathbf U_{n}^{H}$
where
\begin{equation}
\centering
{\mathbf U}_{n}=\left[{\mathbf u}_{P^2+1}, {\mathbf u}_{P^2+2}, \cdots  {{\mathbf u}_{(N_{e}^{s}\times N_{r}^{s} \times N_{f}^{s})^2}}\right]
\end{equation}


Finally, the estimator for the cumulants in the double-array configuration is constructed as follows.
\begin{equation}
\centering
{{P}_{Double\text{ }4-s-MUSICAL}}=\frac{1}{\mathbf d_{4}^{H}{{\mathbf U}_{n}} \mathbf U_{n}^{H}{{\mathbf d}_{4}}}
\end{equation}
with the steering vector ${{\mathbf d}_{4}}=\mathbf d(\theta _{p}^{r},\theta _{p}^{e},{{T}_{p}})\otimes {{\left( \mathbf d(\theta _{p}^{r},\theta _{p}^{e},{{T}_{p}}) \right)}^{*}}$

 \section{Performance test}
In this section, the performance of the proposed algorithm is tested using simulation data in a multi-path environment, small-scale data obtained in an ultrasonic tank \cite{} and ocean data \cite{Roux2008The}, respectively. The test parameters used in these experiments are shown in Table 1 and Table 2.  $z_{s_1}  \sim z_{s_M}$ notes the depths in which the first source to the $M^{th}$ source  is fixed under the water while  $z_{r_1} \sim z_{r_N} $ refers to the ranges which the first receiver to the $N^{th}$ receiver cover. $ D $ is the distance between the reference source and the reference receiver. $ c$ indicates the acoustic velocity in which the raypaths propagate under the water. The central frequency of the emitted signal is $\nu_c$ and its bandwidth is noted as $\nu_w$.  $\nu_n$ notes the number of frequencies used in each test choosing the beginning frequency $\nu_b$ to the last frequency $\nu_e$. $ N_s$ refers to the number of samples used in each test in the time domain.

 Figures 2 and 3 show the separation results of the proposed algorithm while the contrast ones of the smoothing-MUSICAL algorithm and the 4-smoothing-MUSICAL algorithm are also provided.
Specifically, figure 2 (a), (c) and (e) illustrate the separation results of a set of simulation data in the case that five raypaths propagate between the emitter array and the receiver array, which is composed of four receivers. The signal to noise ratio is equal to 2dB. The reference source (receiver) is located at 50m under the water. The smoothing-MUSICAL algorithm provides a two-dimensional (2D) separation in a plan of the arrival time and the direction of arrival (DOA) in figure 2 (a)  while the other two algorithms give a 3D separation with the additional parameter: the direction of emission (DOE) in figure 2 (c) and figure 2 (e), respectively. In figure 2 (a), (c) and (e), each spot corresponds to a raypath and the black crosses mark the theoretical positions. It is obviously that the smoothing-MUSICAL algorithm fails in finding the raypath with the DOA around at  $-5^{\circ}$ in figure 2 (a) and the raypath with the DOA around at  $0^{\circ}$ is deficient in the results of the double-MUSICAL algorithm in figure 2 (c). In contrast, the proposed algorithm successfully separates all the five raypaths without any artifact in figure 2 (e), where the deficient raypaths in figure 2 (a) and figure 2 (c) are highlighted by the red circles.

Figure 2(b), (d) and (f) illustrate the performance contrast of the three algorithms using a set of real data obtained at a small-scale ultrasonic tank. The small-scale experiment reproduces the actual physical phenomena occurring in the real ocean environment in a smaller scale inside the laboratory based on the operation multiplying the frequency of the signals by a factor and dividing the spatial distances by the same factor. Thus, the small-scale experiment is a reduced cost and a totally controlled experiment.
{Specifically, in this tank, a steel bar acts as the bottom, for which the {boundary conditions} are nearly perfect at the water-bottom interface,  and a ~1.10m-long, 5.4-cm-deep acoustic waveguide is constructed {\cite{Roux2014Inverting}}. Two coplanar 64-element vertical line arrays (VLA) are placed and the central frequency of the transducer is 1 or 3 MHz with a $50\%$ frequency bandwidth. The transducer {dimensions} $0.75 mm \times12 mm$ {are} used to make the linear arrays omni-directional in the plane defined by the source-receiver arrays {and the beams collimated in the plane perpendicular to the waveguide axis.}
Figure 2(b), (d) and (f) show a group of separation results with a set of real data obtained in the small-scale experiment. There are seven predictable raypaths in the example, whose theoretical positions are indicated by the black crosses. Figures 2(b) and 1(d) show the separation results using the smoothing-MUSICAL algorithm and the double-MUSICAL algorithm, respectively. It can be clearly seen that both of the two algorithms are deficient in detecting one raypath with the DOA around at $5^{\circ}$. Comparing to them, the proposed 3D higher order algorithm correctly find all the seven raypaths in figure 1(f), where the red circle highlights the deficient raypath in figures 2(b) and 2(d).

Finally, the performance of the double-4-smoothing algorithm is tested using ocean data. The experiment has been performed in July 2005 north of Elba Island, Italy \cite{Roux2008The}. It uses a similar experimental setup to the small-scale ultrasonic experiment, although at a much larger scale. Two equally spaced vertical linear arrays are implemented in 120 m water and the distance between them is 4.071 km. The source array (SA) is composed of 29 transducers covering 78 m and the receiver array (RA) has 32 hydrophones spanning 62m. The central frequency of the transducers is 3.2 kHz with 1 kHz bandwidth.
Figure 3 shows the contrast of separation results using a set of ocean data. There are three expected rays, a surface-reflected ray and two refracted rays. Each refracted ray has a turning point near the depth of maximum sound-speed variability. Due to the sound speed variation described in the Ref.\cite{Roux2008The}, the three raypaths arrived at close time.
Figure 3(a) shows that the separation result of the smoothing-MUSICAL algorithm in the plan of the arrival time and the DOA. It detects just a mixed spot for two of the three raypaths due to their close arrival times and DOAs. Figure 3(b) shows that the separation result of the double-MUSICAL algorithm in the plan of the emitted angle and the DOA, which has a raypath deficiency with the DOA around at $10^{\circ}$.  However, the proposed algorithm gives an accurate separation for the three raypaths shown in the plan of the emitted angle and the DOA in Figure 3 (c) because of the different emitted angles and a strong robustness of the proposed algorithm to the fluctuant ocean. 
Figure 3(d) displays the three raypaths in Figure 3(a)-(c), which propagate between the centers of source and receive arrays. Based on the performance contrasts with the simulation data,  small-scale data and ocean data, a conclusion is drawn that the double-4-smoothing-MUSICAL algorithm enables to achieve an aperture extension, a stronger robustness to the fluctuant environment and a resolution improvement for the extension of cumulants-based algorithm to the double-array configuration.

\begin{table}{}
\begin{center} \begin{tabular}{ c | c | c | c|c| c} \hline & $M$ & $N$ & $z_{s_1} \sim z_{s_M} (m) $ & $z_{r_1} \sim z_{r_N} (m)$ & $d$ (m)\\ \hline \hline Simulation & 4 & 4 &  $47.5 \sim 52.5$& $47.5 \sim 52.5$ & 2.5  \\ \hline \hline Small-scale experiment  & 4&5 & $26.375 \times 10^{-3} \sim 28.625 \times 10^{-3}$ &$25.625 \times 10^{-3} \sim 28.625 \times 10^{-3}$  & $0.75\times 10^{-3}$ \\ \hline\hline Ocean data  & 4&4 & $93.098\sim 101.456 $& 94 $\sim$ 100 & 2  \\ \hline   \end{tabular} \end{center}
\caption{The configuration parameters of simulation, small-scale experiment and the at-sea experiment.}
\end{table}

\begin{table}{}
\begin{center}
\begin{tabular}{ c | c | c|c| c|c |c} \hline & $\nu_c$ (Hz) & $\nu_w$ (Hz) &$ \nu_n (\nu_b \sim\nu_e) $ (Hz) & $ N_s$ & $ D(m)$ & $ c\ (m/s)$
 \\
\hline \hline Simulation & $1.5 \times10^{3}$ & $5 \times10^{3}$ & 75 (0 $\sim$ $5 \times10^{3}$)& 135 & $ 2000 $ & $ 1500$ \\
\hline \hline Small-scale experiment  &1.2 $ \times10^{6}$ & 5 $ \times10^{6}$ & 150(0 $\sim$ 5 $\times10^{6}$)  & 132& $ 1$ & $ 1473$ \\
\hline\hline Ocean data  & $3.2 \times10^{3}$ & $6 \times10^{3}$ & 50 (0 $\sim$ $6 \times10^{3}$) & 60 & $  4.701\times 10^{3}$ & $ 1509$ \\
\hline
\end{tabular}
\end{center}
\caption{{The test parameters used in the simulations, the small-scale experiment and the at-sea experiment.}}
\end{table}

\begin{figure}
\centering     
\subfigure[]{\label{fig:b}\includegraphics[width=62mm]{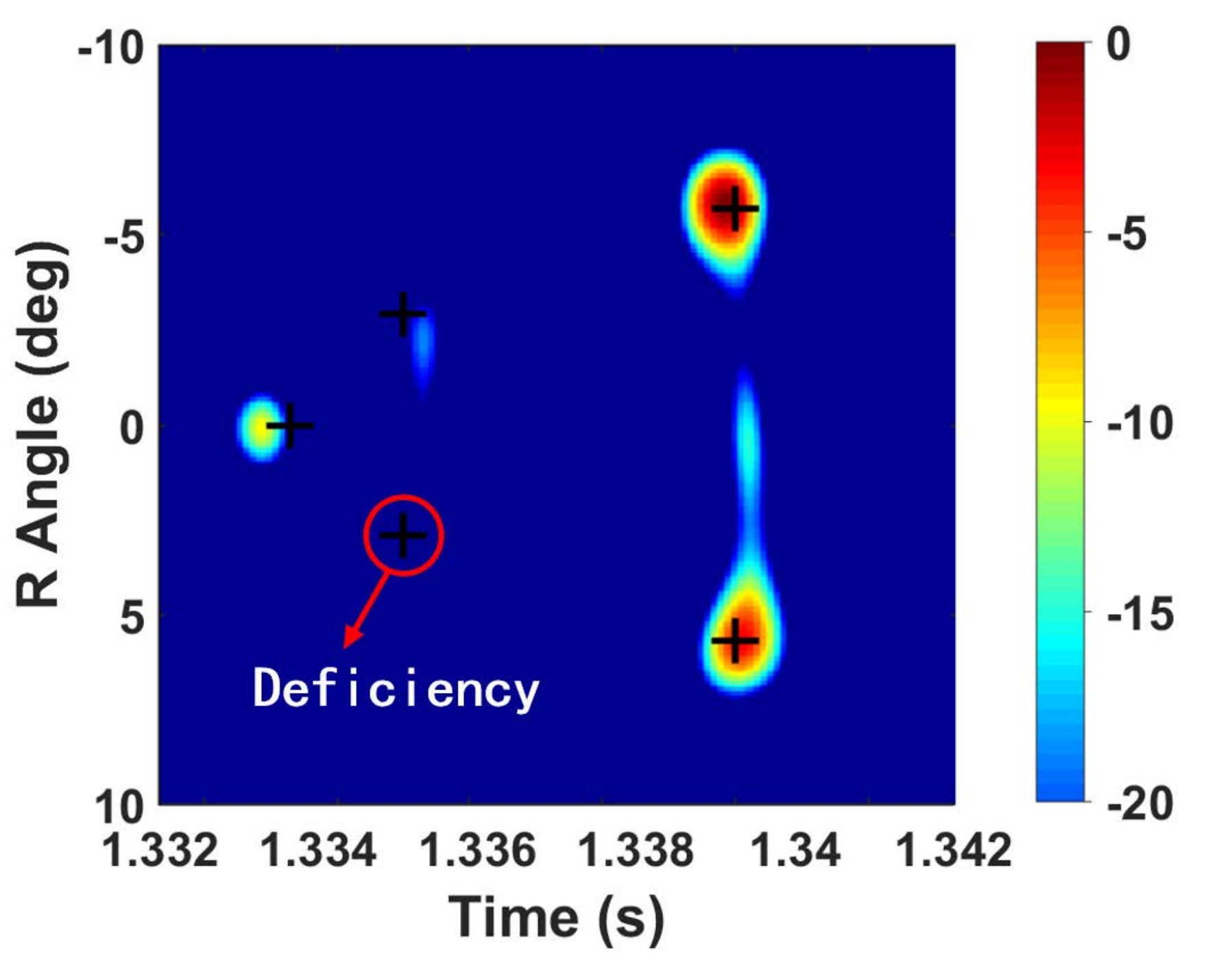}}
\subfigure[]{\label{fig:b}\includegraphics[width=66mm]{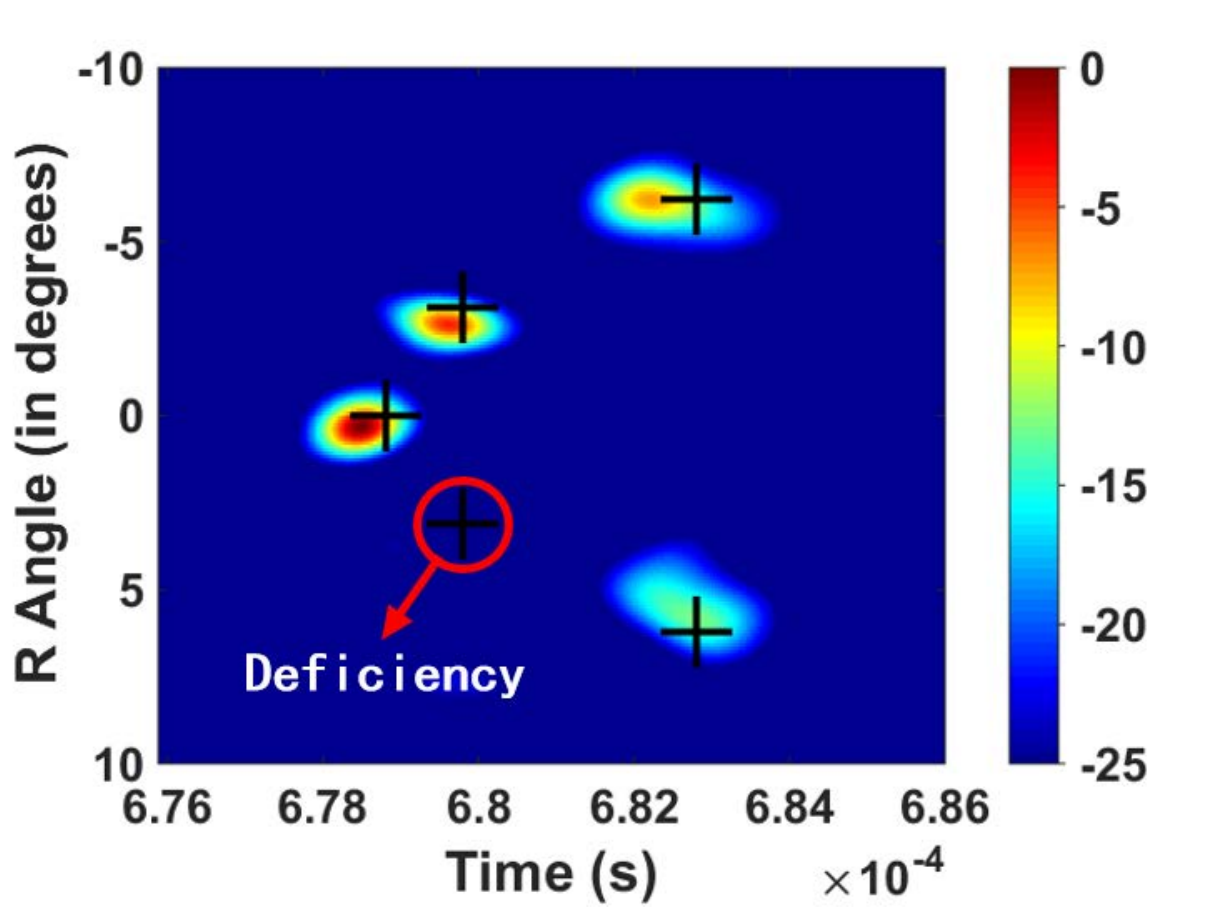}}
\subfigure[]{\label{fig:a}\includegraphics[width=65mm]{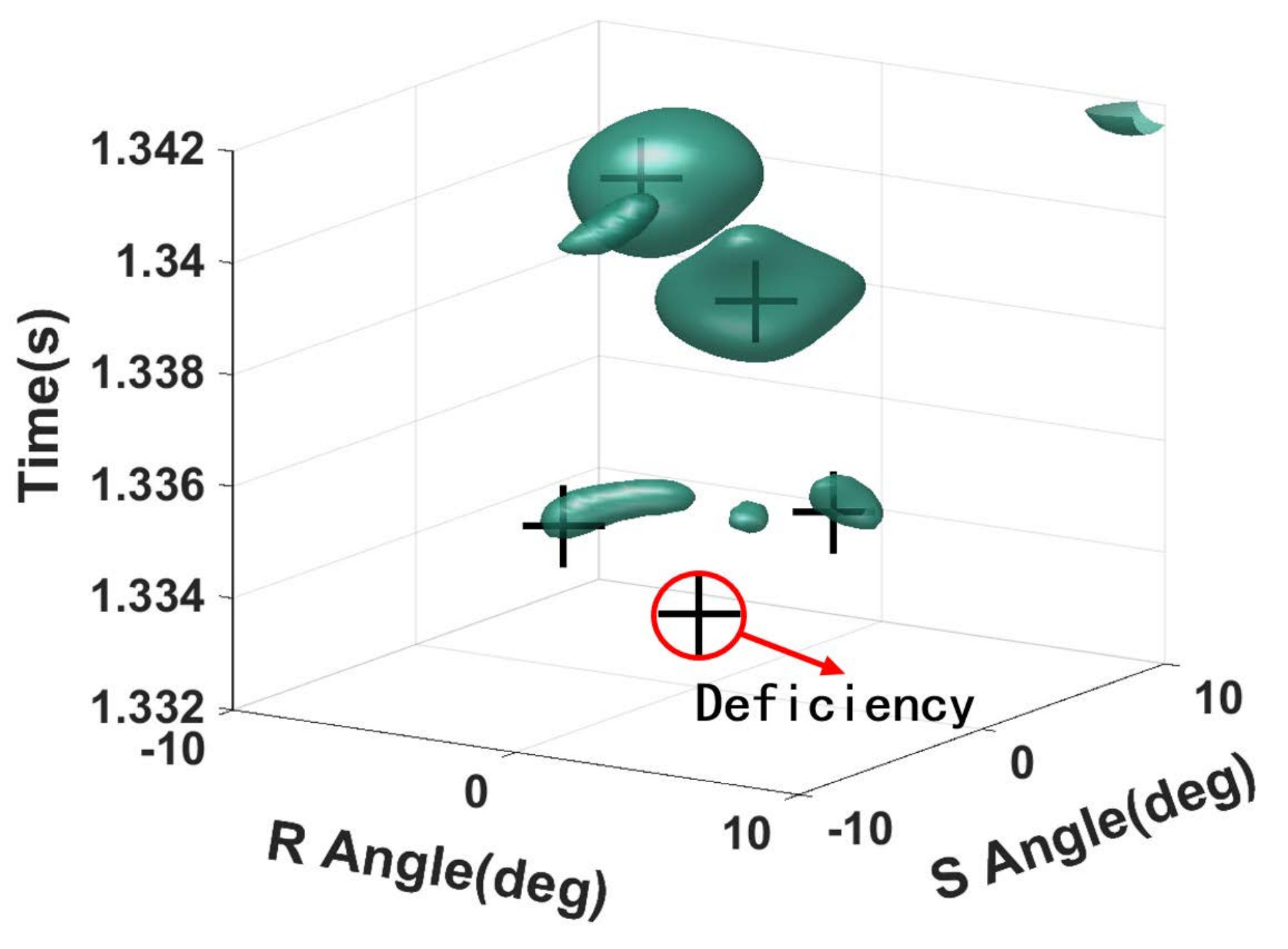}}
\subfigure[]{\label{fig:a}\includegraphics[width=65mm]{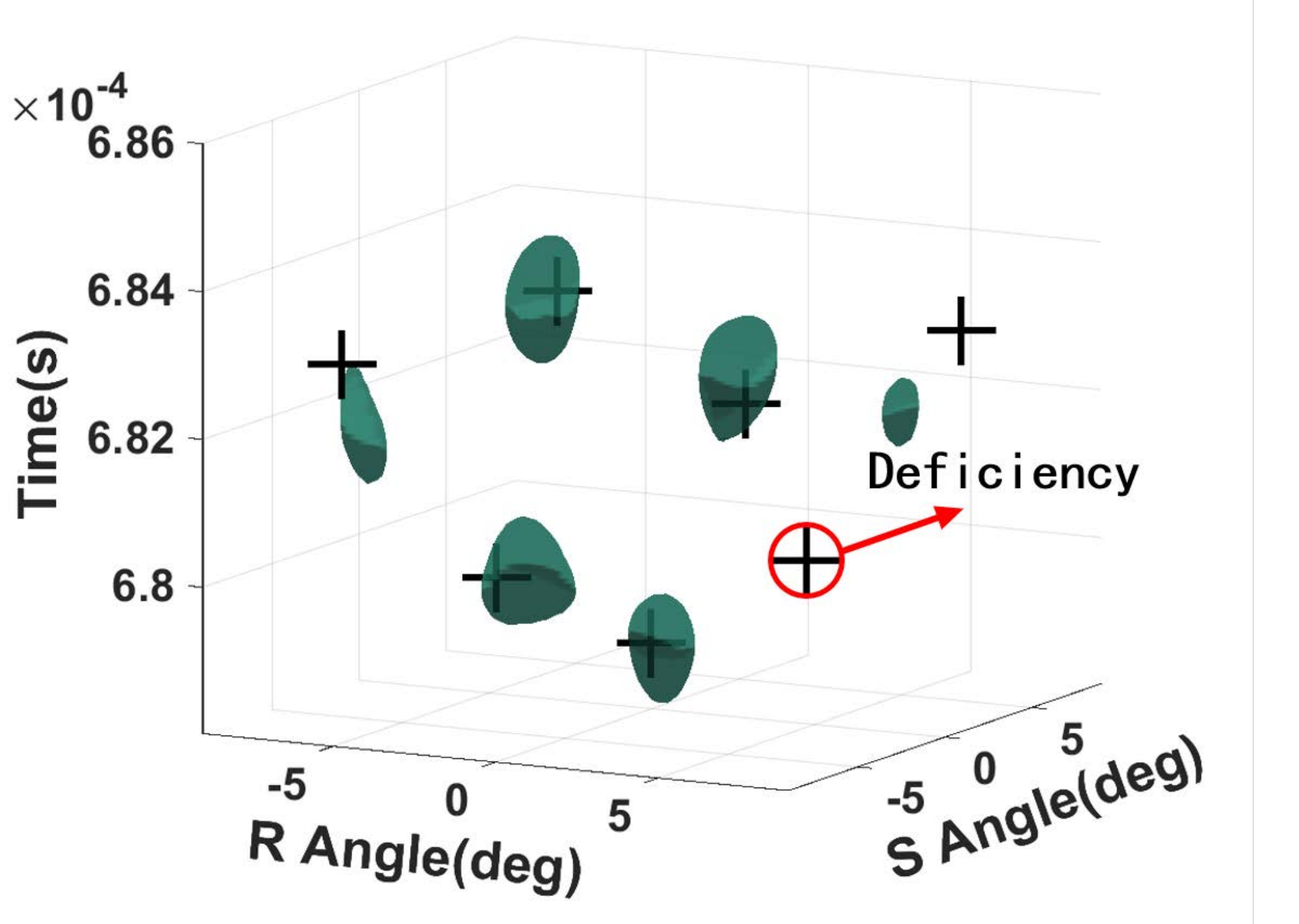}}
\subfigure[]{\label{fig:b}\includegraphics[width=65mm]{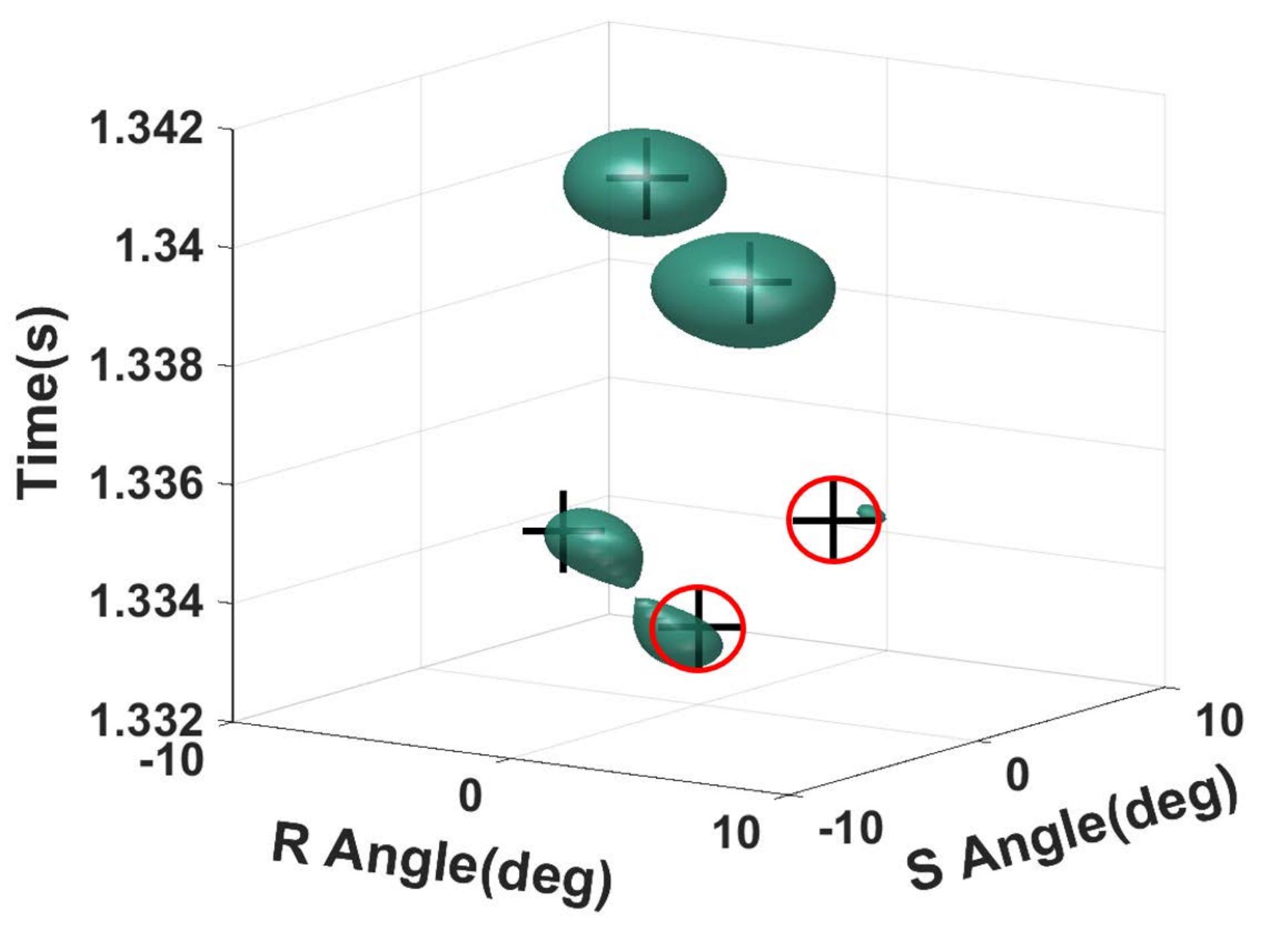}}
\subfigure[]{\label{fig:b}\includegraphics[width=65mm]{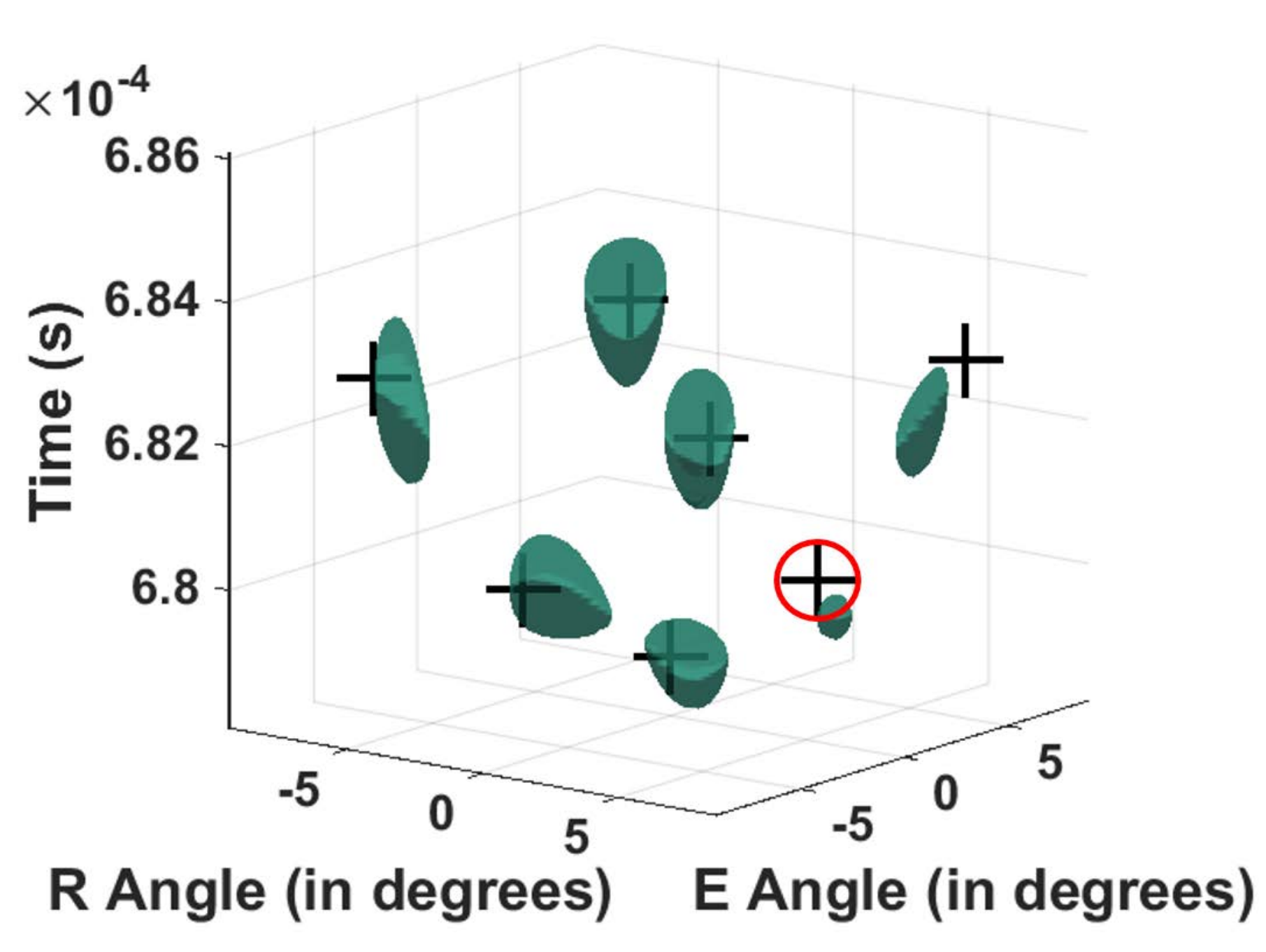}}
\caption{Separation results comparison of different algorithms using simulation datas ((a), (c) and (e)) or using real data obtained in an ultrasonic tank ((b), (d) and (f)). The black crosses denote the theoretical values. The red circles highlight the raypaths used to differentiate the separation ability of the algorithms. (a) and (b) show the results of the smoothing-MUSICAL algorithm; (c) and (d) show the results of the double-MUSICAL algorithm; (e) and (f) show the results of the double-4-MUSICAL algorithm}
\end{figure}


\begin{figure}
\centering     
\subfigure[]{\label{fig:b}\includegraphics[width=76mm, height= 55mm]{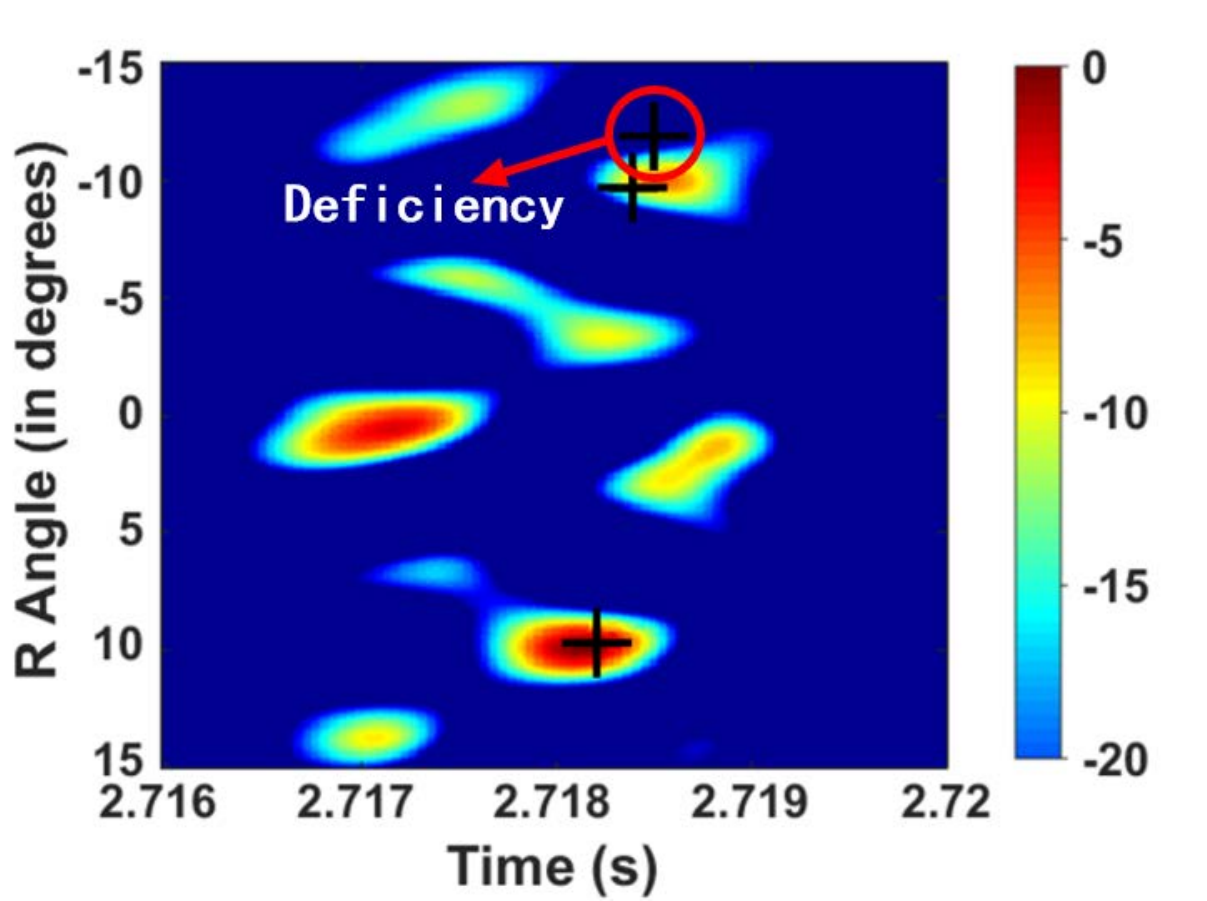}}
\subfigure[]{\label{fig:a}\includegraphics[width=78mm, height= 56mm]{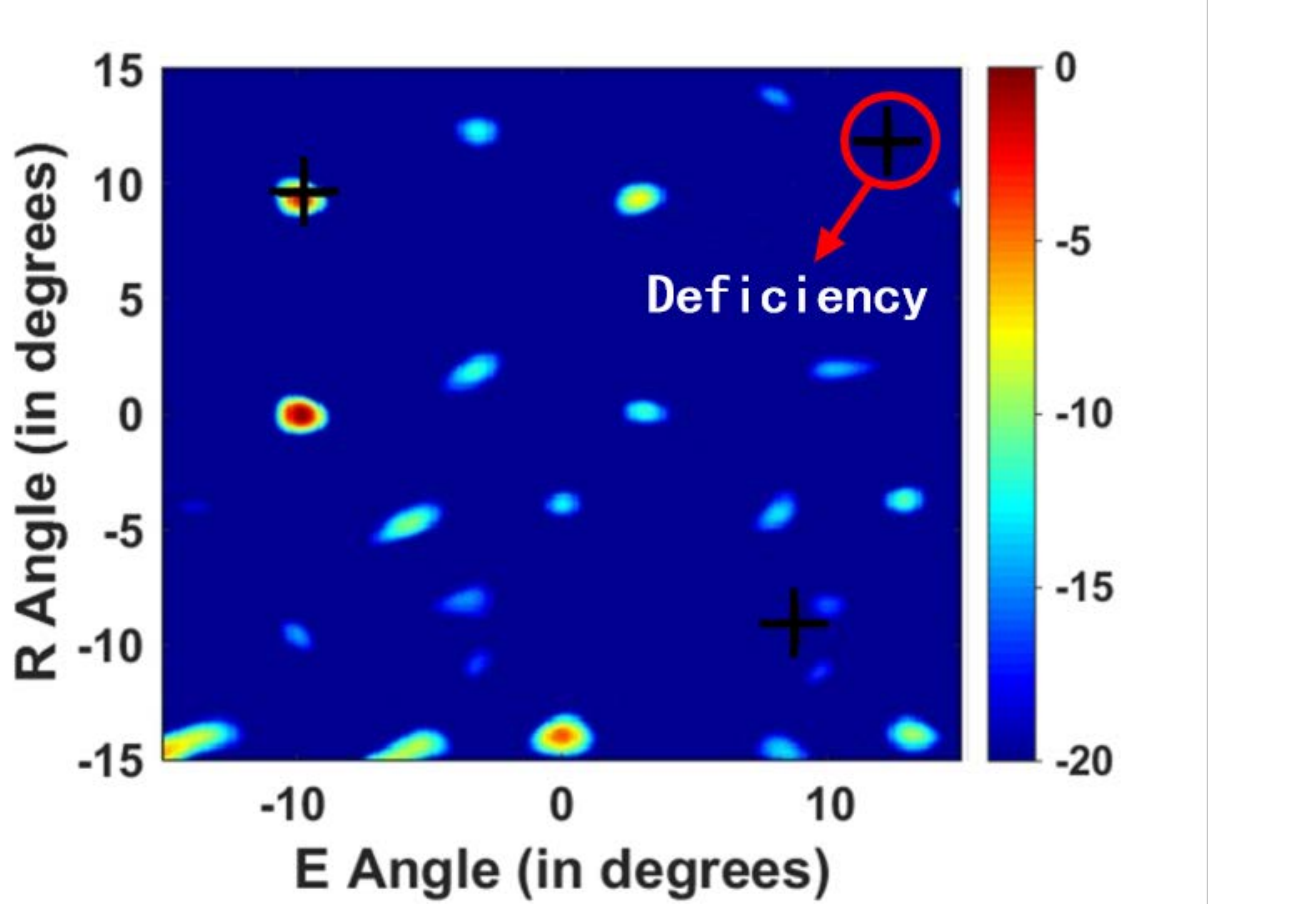}}
\subfigure[]{\label{fig:b}\includegraphics[width=78mm, height= 56mm]{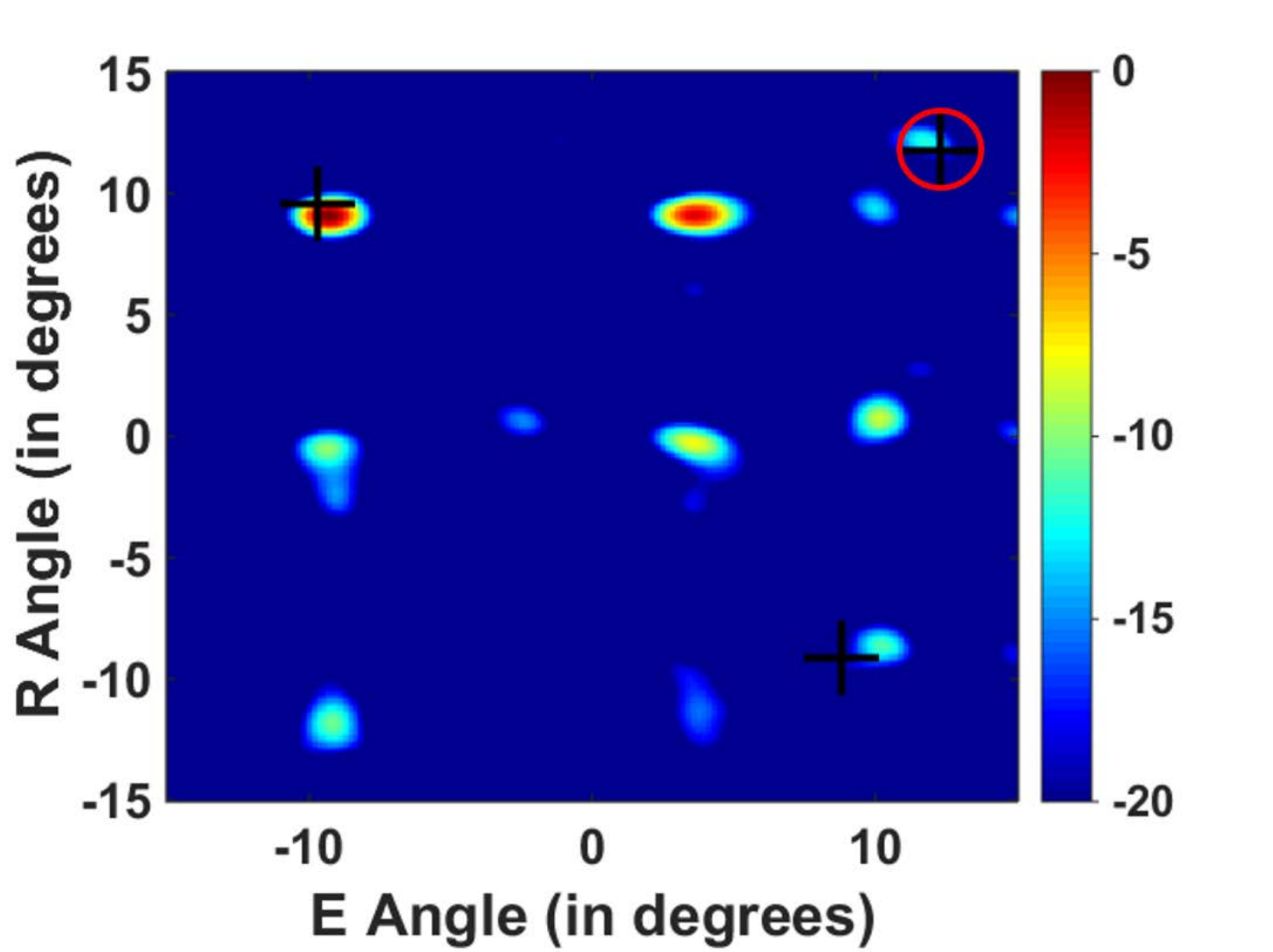}}
\subfigure[]{\label{fig:b}\includegraphics[width=76mm, height= 56mm]{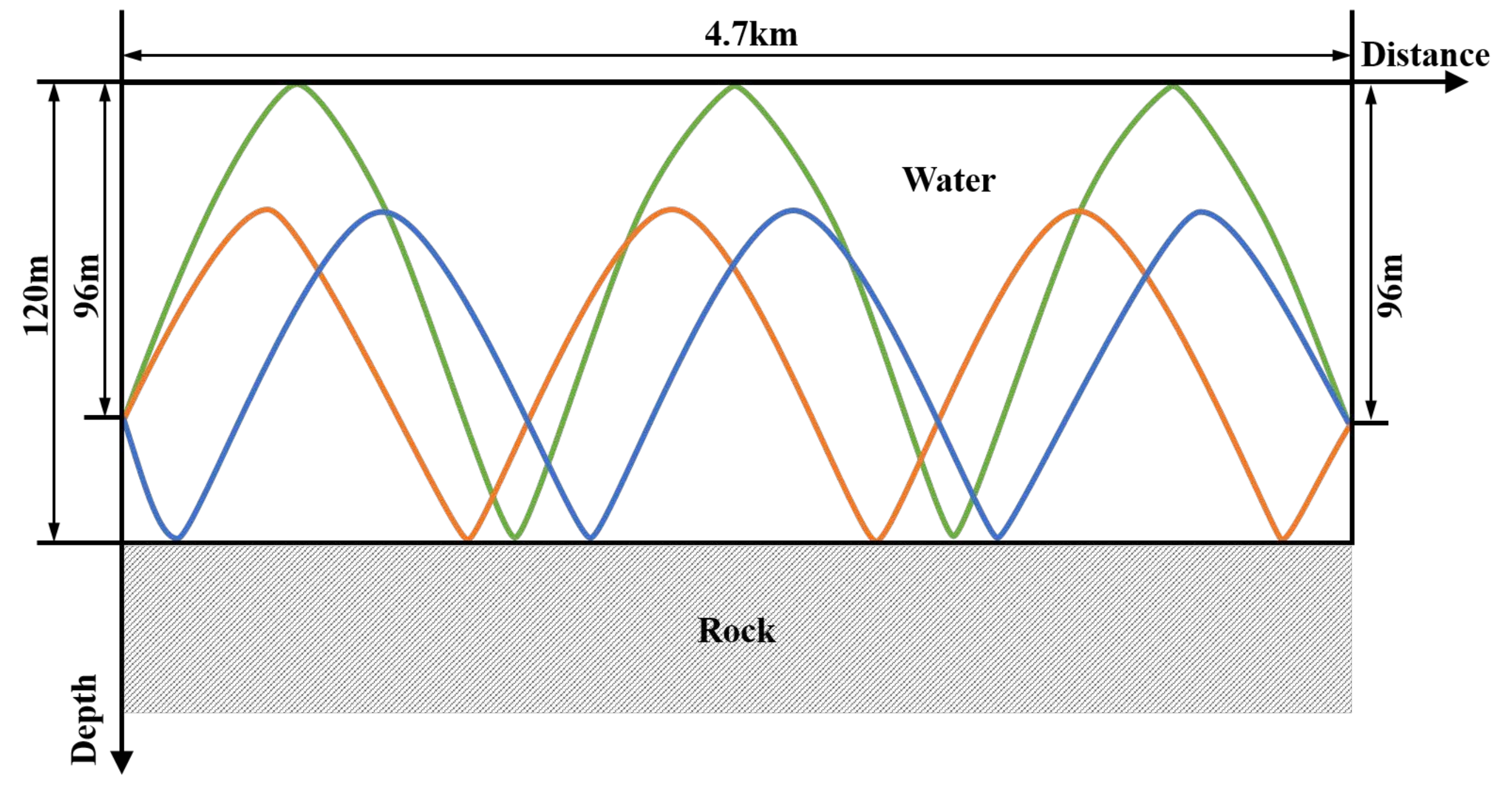}}
\caption{Separation results comparison of different algorithms using ocean datas. The black crosses denote the theoretical values. The red circles highlight the raypaths used to differentiate the separation ability of the algorithms (a) the smoothing-MUSICAL algorithm; (b) the double-MUSICAL algorithm; (c) the double-4-MUSICAL algorithm; (d) Raypaths between the source and the reference sensor for the three rays.}
\end{figure}


\section{Conclusion}
In this paper, we extended a higher-order algorithm to a double-array configuration. Higher-resolution and more robust separation for close arrivals even in a fluctuant ocean environment is achieved by the 3D algorithm. In future work, reducing its computation complexity will be considered.


\begin{acknowledgments}
This resarch has been supported by the National Natural Science Foundation of China (Nos. 61871124 and 61876037), the State Key Laboratory of Acoustics, Chinese Academy of Sciences (No. SKLA201604), and the Scientific Research Foundation for the Returned Overseas Chinese Scholars. The Focused Acoustic Forecasting experiment (FAF05) experiment was performed in collaborative experiments with the NATO Underwater Research Centre (NURC), La Spezia, Italy, with Mark Stevenson as Chief Scientist. Scientists who contributed to these experiments include Tuncay Akal, W. A. Kuperman, W. H. Hodgkiss, H.C. Song, B.D. Cornuelle, Piero Boni, Piero Guerrini, other NURC staff, and the officers and crew of the RV Alliance.
\end{acknowledgments}







\end{document}